\documentclass[italian,english]{article}
\usepackage[latin9]{inputenc}
\usepackage{color}
\usepackage{amssymb}

\newcommand{\lyxaddress}[1]{
\par {\raggedright #1
\vspace{1.4em}
\noindent\par}
}

\usepackage{babel}

\begin{document}

\title{\textbf{Inflation from $R^{2}$ gravity: a new approach using nonlinear
electrodynamics}}

\author{\textbf{$^{+}$Christian Corda and $^{*}$Herman J Mosquera Cuesta}}

\maketitle

\lyxaddress{\begin{center}
\textbf{$^{+}$}Institute for Basic Research, P. O. Box 1577, Palm
Harbor, FL 34682, USA %
\footnote{\begin{quotation}
C. Corda is partially supported by a Research Grant of The R. M. Santilli
Foundations Number RMS-TH-5735A2310
\end{quotation}
}and Associazione Galileo Galilei, Via B. Buozzi 47, 59100 PRATO, Italy
\par\end{center}}

\begin{center}
\textbf{$^{*}$}Instituto de Cosmologia, Relatividade e Astrofìsica
(ICRA-BR), Centro Brasilero de Pesquisas Fisicas, Rua Dr. Xavier Sigaud
150, CEP 22290 - 180 Urca Rio de Janeiro - RJ Brazil
\par\end{center}

\lyxaddress{\begin{center}
\textit{E-mail addresses:}\textbf{$^{+}$}\textcolor{blue}{cordac.galilei@gmail.com;
}\textbf{$^{*}$}\textcolor{blue}{herman@icra.it;}
\par\end{center}}
\begin{abstract}
We discuss another approach regarding the inflation from the \textbf{$R^{2}$}
theory of gravity originally proposed by Starobinski. A non-singular
early cosmology is proposed, where, adding a nonlinear electrodynamics
Lagrangian to the high-order action, a bouncing is present and a power-law
inflation is obtained. In the model the Ricci scalar $R$ works like
an inflaton field.
\end{abstract}

\lyxaddress{PACS numbers: 04.50.+h, 04.20.Jb.}


\section{Introduction}

The accelerated expansion of the Universe that is currently purported
from observations of SNe Ia suggests that cosmological dynamics is
dominated by a ``new'' substance of the universe constituents dubbed
as Dark Energy, which is able to provide a large negative pressure
to account for the late-time accelerate expansion. This is the standard
picture, in which such a new ingredient is considered as a source
of the \emph{right-hand-side} of the field equations. It is posed
that it should be some form of un-clustered non-zero vacuum energy
which, together with the clustered Dark Matter, drives the global
dynamics. This is the so-called {}``concordance model''($\Lambda$CDM)
which gives, in agreement with the data analysis of the observations
of the Cosmic Microwave Background Radiation (CMBR), Lyman Limit Systems
(LLS) and type la supernovae (SNe Ia), a good framework for understanding
the currently observed Universe. However, the $\Lambda$CDM presents
several shortcomings as the well known``coincidence'' and ``cosmological
constant'' problems \cite{key-1}. 

An alternative approach to explain the purported late-time acceleration
of the universe is to change the \textit{left hand side} of the field
equations, and to inquire whether the observed cosmic dynamics can
be achieved by extending general relativity \cite{key-2,key-3,key-4}.
In this different context, it is not required to search candidates
for Dark Energy and Dark Matter, which until to date, have not been
found, but rather it claims that only the {}``observed'' ingredients:
curvature and baryon matter, have to be taken into account. Considering
this point of view, one can posit that gravity is not scale-invariant
\cite{key-5}. In so doing, one allows for a room for alternative
theories to be opened \cite{key-6,key-7,key-8}. In principle, interesting
Dark Energy and Dark Matter models can be built by considering $f(R)$
theories of gravity \cite{key-5,key-9} (here $R$ is the Ricci curvature
scalar). 

In this perspective, even the sensitive detectors of gravitational
waves like bars and interferometers (i.e., those which are currently
in operation and the ones which are in a phase of planning and proposal
stages\cite{key-10,key-11}, could, in principle, test the physical
consistency of general relativity or of any other theory of gravitation.
This is because in the context of Extended Theories of Gravity important
differences with respect to general relativity show up after studying
the linearized theory\cite{key-12,key-13,key-14,key-15}.

In this paper, another approach regarding the inflation from the \textbf{$R^{2}$}
theory of gravity, which is the simplest among $f(R)$ theories and
was been originally proposed by Starobinski in \cite{key-16}, is
shown. A non-singular early cosmology is proposed, where, adding a
nonlinear electrodynamics Lagrangian to the high-order action, a bouncing
is present and a power-law inflation is obtained. In the model the
Ricci scalar $R$ works like an inflaton field.

In the general picture of high order theories of gravity, recently
the \textbf{$R^{2}$} theory has been analysed in various interesting
frameworks, see \cite{key-17,key-18} for example. 

We recall that extensions of the traditional Maxwell electromagnetic
Lagrangian, which take into account high order terms of the electromagnetic
scalar F, have been used in cosmological models \cite{key-19}, gravitational
redshifts of neutron stars \cite{key-20} and pulsars \cite{key-21}.
Moreover, a particular nonlinear Lagrangian has been analysed in the
context of the Pioneer 10/11 spacecraft anomaly \cite{key-22}.

\section{Action and Lagrangian}

Let us consider the high order action \cite{key-16,key-17,key-18}
\begin{equation}
S=\int d^{4}x\sqrt{-g}(R+\alpha R^{2}+\mathcal{L}_{m}).\label{eq: high order 1}\end{equation}

Such an equation (\ref{eq: high order 1}) is a particular choice
in respect to the well known canonical one of General Relativity (the
Einstein - Hilbert action \cite{key-23}) which is 

\begin{equation}
S=\int d^{4}x\sqrt{-g}(R+\mathcal{L}_{m}).\label{eq: EH}\end{equation}

We are going to show that the action (\ref{eq: high order 1}), applied
to the Friedman-Robertson-Walker Cosmology, generates a non singular
inflationary phase of the Universe where the Ricci scalar acts like
inflaton, and a bouncing is present,  if $\mathcal{L}_{m}$ is the
non linear electrodynamics Lagrangian. Note that in this letter we
work with $8\pi G=1$, $c=1$ and $\hbar=1.$

Inflationary models of the early Universe were analysed in the early
and middles 1980's (see \cite{key-24} for a review), starting from
an idea of Starobinski \cite{key-15} and Guth \cite{key-25}. These
are cosmological models in which the Universe undergoes a brief phase
of a very rapid expansion in early times. In this context the expansion
could be power-law or exponential in time. Inflationary models provide
solutions to the horizon and flatness problems and contain a mechanism
which creates perturbations in all fields \cite{key-24}. 

In Cosmology, the Universe is seen like a dynamic and thermodynamic
system in which test masses (i.e. the {}``particles'') are the galaxies
that are stellar systems with a number of the order of $10^{9}-10^{11}$
stars \cite{key-23}. Galaxies are located in clusters and super clusters,
and observations show that, on cosmological scales, their distribution
is uniform. This is also confirmed by the WMAP data on the Cosmic
Background Radiation \cite{key-26,key-27}. These assumption can be
summarized in the so called Cosmological Principle: \textit{the Universe
is homogeneous everywhere and isotropic around every point.} Cosmological
Principle simplifies the analysis of the large scale structure, because
it implies that the proper distances between any two galaxies is given
by an universal scale factor which is the same for any couple of galaxies
\cite{key-23}.

In this framework, the cosmological line - element is the well known
Friedman-Robertson-Walker one, and for a sake of simplicity we will
consider the flat case, because the WMAP data are in agreement with
it \cite{key-26,key-27}:\begin{equation}
ds^{2}=-dt^{2}+a^{2}(dz^{2}+dx^{2}+dy^{2}).\label{eq: metrica FRW}\end{equation}
Following \cite{key-23} we also get \begin{equation}
g_{\mu\nu}=\begin{array}{cccc}
-1 & 0 & 0 & 0\\
0 & +a^{2} & 0 & 0\\
0 & 0 & +a^{2} & 0\\
0 & 0 & 0 & +a^{2},\end{array}\label{eq: g}\end{equation}

\begin{equation}
\sqrt{-g}=a^{3}\label{eq: radg}\end{equation}

and \begin{equation}
R=-6[\frac{1}{a}\frac{d\dot{a}}{dt}+(\frac{\dot{a}}{a})^{2}].\label{eq: Ricci Scalar}\end{equation}

One can use the Lagrange multipliers putting

\begin{equation}
S=2\pi^{2}\int dt\{a^{3}(R+\alpha R^{2})-\beta[R+6\frac{\ddot{a}}{a}+6(\frac{\dot{a}}{a})^{2}]+a^{3}\mathcal{L}_{m}\}.\label{eq: high order felix}\end{equation}

$\beta$ can be obtained by varing the action in respect to $R$.
It is

\begin{equation}
a^{3}\frac{\partial(R+\alpha R^{2})}{\partial R}\delta R-\beta\delta R,\label{eq: Lagrange}\end{equation}

which gives

\begin{equation}
\beta=a^{3}\frac{\partial(R+\alpha R^{2})}{\partial R}=a^{3}(2\alpha R+1).\label{eq: Lagrange 2}\end{equation}

Thus, substituting in eq. (\ref{eq: high order felix}) one obtains

\begin{equation}
S=2\pi^{2}\int dt\{-2a^{3}\alpha R^{2}-6a^{2}\ddot{a}(2\alpha R+1)-6a(\dot{a})^{2}(2\alpha R+1)+a^{3}\mathcal{L}_{m}\}.\label{eq: high order felix 2}\end{equation}

The term $-6a^{2}\ddot{a}(2\alpha R+1)$ is critical as it contains
a second derivative of $a$. Let us integrate it. It is

\begin{equation}
\begin{array}{c}
-6\int dta^{2}\ddot{a}(2\alpha R+1)=-6a^{2}\dot{a}(2\alpha R+1)+6\int dt[2\alpha a^{2}\dot{a}\dot{R}+2a(\dot{a})^{2}(2\alpha R+1)]=\\
\\=6\int dt[2\alpha a^{2}\dot{a}\dot{R}+2a(\dot{a})^{2}(2\alpha R+1)],\end{array}\label{eq: int parti}\end{equation}

where we have taken into account that the term outside the integral
is equal to zero as it is a pure divergence.

Substituting in eq. (\ref{eq: high order felix 2}), one gets

\begin{equation}
S=2\pi^{2}\int dt\{-a^{3}\alpha R^{2}+12\alpha a^{2}\dot{a}\dot{R}+6a(\dot{a})^{2}(2\alpha R+1)+a^{3}\mathcal{L}_{m}\}.\label{eq: high order felix 3}\end{equation}

Then, the Lagrangian is \begin{equation}
\mathcal{L}=-a^{3}\alpha R^{2}+12\alpha a^{2}\dot{a}\dot{R}+6a(\dot{a})^{2}(2\alpha R+1)+a^{3}\mathcal{L}_{m}.\label{eq: RLC 4}\end{equation}

The energy function associated to the Lagrangian is \cite{key-23}\begin{equation}
E_{\mathcal{L}}=\frac{\partial\mathcal{L}}{\partial\dot{a}}\dot{a}+\frac{\partial\mathcal{L}}{\partial\dot{R}}\dot{R}-\mathcal{L}.\label{eq: en lagr}\end{equation}

Combining eq. (\ref{eq: RLC 4}) with eq. (\ref{eq: en lagr}), the
condition \begin{equation}
E_{\mathcal{L}}=0\label{eq: condizione}\end{equation}
 together with the definition of the Hubble constant, i.e. $H=\frac{1}{a}\frac{da}{dt},$
and with a little algebra gives

\begin{equation}
H^{2}=\frac{\mathcal{L}_{m}}{3\alpha R}-H\frac{\dot{R}}{R}.\label{eq: Hubble}\end{equation}

From the Eulero-Lagrange equation for $a$ and $\dot{a},$ i.e. \cite{key-23}

\begin{equation}
\frac{\partial\mathcal{L}}{\partial a}=\frac{d}{dt}(\frac{\partial\mathcal{L}}{\partial\dot{a}})\label{eq: el}\end{equation}

one gets

\begin{equation}
\ddot{R}+3H\dot{R}=\frac{2\mathcal{L}_{m}}{3\alpha}.\label{eq: Energy}\end{equation}

An important question is where Eq. (15) comes from \cite{key-29}.
In general relativity, due to the reparametrization invariance of
the time coordinate, the total energy (including the contribution
from the gravity sector) vanishes \cite{key-29}. In the action (12),
however, there is not the reparametrization invariance because the
total derivative terms are dropped \cite{key-29}. Then, one can think
that the total energy does not always vanish \cite{key-29}. We clarify
this point as it follows. Let us start by the original action (\ref{eq: high order 1})
from which the action (12) arises. Let us consider the conformal transformation
\cite{key-30} \begin{equation}
\tilde{g}_{\alpha\beta}=e^{2\Phi}g_{\alpha\beta}\label{eq: conforme}\end{equation}

where the conformal rescaling \begin{equation}
e^{2\Phi}=2\alpha R+1\label{eq: rescaling}\end{equation}

has been chosen. By applying the conformal transformation (\ref{eq: conforme})
to the action (\ref{eq: high order 1}) the conformal equivalent Hilbert-Einstein
action \begin{equation}
A=d^{4}x\sqrt{-\widetilde{g}}[\widetilde{R}+\mathcal{L}(\Phi,\Phi_{;\alpha})+\mathcal{L}_{m}],\label{eq: conform}\end{equation}

is obtained. $\mathcal{L}(\Phi,\Phi_{;\alpha})$ is the conformal
scalar field contribution derived from

\begin{equation}
\tilde{R}_{\alpha\beta}=R_{\alpha\beta}+2(\Phi_{;\alpha}\Phi_{;\beta}-g_{\alpha\beta}\Phi_{;\delta}\Phi^{;\delta}-\frac{1}{2}g_{\alpha\beta}\Phi^{;\delta}{}_{;\delta})\label{eq: conformRicci}\end{equation}

and \begin{equation}
\tilde{R}=e^{-2\Phi}+(R-6\square\Phi-6\Phi_{;\delta}\Phi^{;\delta}).\label{eq: conformRicciScalar}\end{equation}

Clearly, the reparametrization invariance of the time coordinate is
consistent with the new action (\ref{eq: conform}) in the conformal
Einstein frame and the total energy (including the contribution from
the gravity sector) vanishes in this case too. One could object that
the energy in the conformal Einstein frame is different with respect
to the energy in the original Jordan frame, but in ref. \cite{key-31}
it has been shown that the two conformal frames are energetically
equivalent if, together with the conformal rescaling (\ref{eq: conforme}),
times and lengths are rescaled as $e^{\Phi}$ while the mass-energy
is rescaled as $e^{-\Phi}.$ This analysis permits to enable the condition
of Eq. (15) in the present discussion too.

\section{Nonlinear electrodynamics Lagrangian and Inflation}

In order to show that our model admits a power law inflationary phase,
we need to postulate some matter Lagrangian $\mathcal{L}_{m}$ which
can perform the condition of inflation $P<-\rho$ \cite{key-24}.
We will use the non linear electrodynamics Lagrangian of \cite{key-19},
which is \begin{equation}
\mathcal{L}_{m}\equiv-\frac{1}{4}F+c_{1}F^{2}+c_{2}G^{2},\label{eq: NLD}\end{equation}

where $F$ is the electromagnetic scalar, $c_{1},\textrm{ }\textrm{ }c_{2}$
are two constants and, considering the electromagnetic field tensor
$F^{\alpha\beta}$ (see \cite{key-23} the definition of this object),
$G$ is defined like \cite{key-19} $G\equiv\frac{1}{2}\eta_{\alpha\beta\mu\nu}F^{\alpha\beta}F^{\mu\nu}.$ 

The Lagrangian (\ref{eq: NLD}), differently from the one of the singular
Einstein-Maxwell Universe, performs a non singular Universe with \textit{bouncing}
\cite{key-19}\textit{.} This is because the energy condition of singularity
theorems \cite{key-28} is not satisfied in the case of the non linear
electrodynamics Lagrangian (see \cite{key-19} for details).

In fact, following \cite{key-19}, one uses the equation of state

\begin{equation}
p=\frac{1}{3}\rho-\rho_{*},\label{eq: star}\end{equation}

where 

\begin{equation}
\rho_{*}\equiv\frac{16}{3}c_{1}B^{4}\label{eq: def star}\end{equation}

(see eq. 15, 16 and 25 of \cite{key-19}) and $B$ is the magnetic
field associed to $F.$ 

This equation of state is no longer given by the Maxwellian value,
thus, using eq. (\ref{eq: NLD}), from eqs. (\ref{eq: Hubble}) and
(\ref{eq: Energy}) one gets \begin{equation}
B=\frac{B_{0}}{2a^{2}},\label{eq: B}\end{equation}

where $B_{0}$ is a constant \cite{key-19}, and 

\begin{equation}
\dot{a}^{2}=\frac{B_{0}^{2}}{12\alpha a^{2}R}(1-\frac{8c_{1}B_{0}^{2}}{a^{4}})-2Ha^{2}\frac{\dot{R}}{R},\label{eq: Hubble 2}\end{equation}

which can be solved by suitably choosing the origin of time. 

One gets

\begin{equation}
a^{2}=\frac{B_{0}}{\sqrt{\alpha}}\sqrt{\frac{2}{3}(t^{2}+12c_{1})}.\label{eq: soluzione}\end{equation}

This expression is not singular for $c_{1}>0.$ In this case we see
that at the instant $t=0$ a minimum value of the scale factor is
present:

\begin{equation}
a_{min}^{2}=\frac{B_{0}}{\sqrt{\alpha}}\sqrt{8c_{1}}.\label{eq: minimo}\end{equation}

This also impies that, for a value $t=\sqrt{12c_{1}}$, the energy
density $-\rho$ reaches a maximum value $\rho_{max}=1/64c_{1}.$
For smaller values of $t$ the energy density decreases, vanishing
a $t=0$, while the pressure becomes negative \cite{key-19}.

In this way, the condition of inflation $P-\rho_{*}<0$ \cite{key-24}
gives the inflationary solutions for equations (\ref{eq: Hubble})
and (\ref{eq: Energy}), if one assumes that the Ricci scalar $R$
acts like inflaton:

\begin{equation}
\begin{array}{c}
R(t)\simeq(1+Ht/\beta)^{2}\\
\\a_{inf}(t)\simeq(1+Ht/\beta)^{w+1/2},\end{array}\label{eq: Inflation}\end{equation}

with $\beta\simeq w$ and \begin{equation}
H_{inf}\simeq\sqrt{\mathcal{L}_{m}^{*}},\label{eq: Hubby}\end{equation}

where $\mathcal{L}_{m}^{*}$ is the right hand side of equation (\ref{eq: Hubble})
which is constant during the inflationary phase. The idea of considering
the Ricci scalar as an effective scalar field (scalaron) arises from
Starobinski \cite{key-16}.

\section{Conclusion remarks}

Another approach regarding the inflation from the \textbf{$R^{2}$}
theory of gravity, which was originally proposed by Starobinsk, has
been analysed. A non-singular early cosmology has been proposed, where,
adding a nonlinear electrodynamics Lagrangian to the high-order action,
a bouncing is present and a power-law inflation is obtained. In the
model which has been discussed, the Ricci scalar $R$ works like an
inflaton field.

\section*{Acknowledgements }

We would like to thank Professor Mario Novello for useful discussions
on the topics of this paper. We also thank an unknown referee for
precious advices and suggestions which permitted to improve this paper.

\end{document}